\documentclass[aps,pra,superscriptaddress,twocolumn,longbibliography]{revtex4-2}

\usepackage{graphicx}
\usepackage{amsmath} 
\usepackage{amssymb}
\usepackage{hyperref}
\usepackage[utf8]{inputenc}
\usepackage{mathtools}
\usepackage[english]{babel}
\usepackage{bbm}
\hypersetup{colorlinks=true, linkcolor=blue, citecolor=blue, urlcolor=blue}
\usepackage{xcolor}
\usepackage{braket}
\usepackage{array}
\usepackage{soul}
\newcolumntype{P}[1]{>{\centering\arraybackslash}p{#1}}

\newcommand{\ie}{i.\,e.,\ }

\newcommand{\cf}{cf.\ }
\newcommand{\re}{\mathrm{Re}}
\newcommand{\im}{\mathrm{Im}}

\newcommand{\rr}{\mathbf{r}}

\newcommand{\fref}[1]{\text{Fig.}~\ref{#1}}

\newcommand{\eref}[1]{\text{Eq.}~\eqref{#1}}

\begin{document}
\title{Optimized geometries for cooperative photon storage in an impurity coupled to a two-dimensional atomic array}
\author{Samuel Buckley-Bonanno}
\affiliation{Department of Physics, Harvard University, Cambridge, Massachusetts 02138, USA}
\author{Stefan Ostermann}
\affiliation{Department of Physics, Harvard University, Cambridge, Massachusetts 02138, USA}
\author{Oriol Rubies-Bigorda}
\affiliation{Physics Department, Massachusetts Institute of Technology, Cambridge, Massachusetts 02139, USA}
\affiliation{Department of Physics, Harvard University, Cambridge, Massachusetts 02138, USA}
\author{Taylor L. Patti}
\affiliation{Department of Physics, Harvard University, Cambridge, Massachusetts 02138, USA}
\affiliation{NVIDIA, Santa Clara, California 95051, USA}
\author{Susanne F. Yelin}
\email{syelin@g.harvard.edu}
\affiliation{Department of Physics, Harvard University, Cambridge, Massachusetts 02138, USA}

\begin{abstract} 
The collective modes of two-dimensional ordered atomic arrays can modify the radiative environment of embedded atomic impurities. We analyze the role of the lattice geometry on the impurity's emission linewidth by comparing the effective impurity decay rate obtained for all non-centered Bravais lattices and an additional honeycomb lattice. We demonstrate that the lattice geometry plays a crucial role in determining the effective decay rate for the impurity. In particular, we find that the minimal effective decay rate appears in lattices where the number of the impurity's nearest neighbors is maximal \emph{and} the number of distinct distances among nearest neighbors is minimal. We further show that, in the choice between interstitial and substitutional placement of the impurity, the former always wins by exhibiting a lower decay rate and longer photon storage. For interstitial placements, we determine the optimal impurity position in the lattice plane, which is \emph{not} necessarily found in the center of the lattice plaquette. 
\end{abstract}

\maketitle

\section{Introduction}
Light-matter quantum interfaces~\cite{hammerer_quantum_2010} are a crucial building block for future quantum technologies.
They are a necessity for building up networks of quantum information processors~\cite{kimble_quantum_2008,wehner_quantum_2018}, where an efficient link between photonic degrees of freedom and atoms or other solid-state based quantum processors is decisive. A broad variety of potential platforms realizing efficient light-matter interfaces are currently explored theoretically and experimentally. Prominent examples are single atoms or ions in cavities~\cite{wilk_single-atom_2007, casabone_enhanced_2015}, quantum dots~\cite{lodahl_interfacing_2015, lodahl_quantum-dot_2018} or excitons in two-dimensional solid-state materials~\cite{Palacios-Berraquero2018,back_realization_2018,andersen_beam_2022}.

Recently, arrays of quantum emitters were found to be a versatile tool to enhance and control the interaction between single photons and quantum matter~\cite{chang_cavity_2012,mirhosseini_cavity_2019,masson_atomic-waveguide_2020,masson_many-body_2020,bettles_enhanced_2016,asenjo-garcia_exponential_2017,shahmoon_cooperative_2017,bekenstein_quantum_2020,rui_subradiant_2020,solntsev_metasurfaces_2021,fernandez-fernandez_tunable_2022}. If the interatomic distance is smaller than the atomic transition wavelength, these arrays exhibit cooperative effects due to light-induced resonant dipole-dipole interactions among the single emitters~\cite{reitz_cooperative_2022}. This enhances the effective cross-section of the array, as impinging photons excite collective lattice modes~\cite{perczel_photonic_2017}. The precise control over these collective lattice modes results in a broad variety of potential applications in future quantum technologies. Some examples are the efficient storage and retrieval of photons by dynamically populating subradiant lattice modes~\cite{rubies-bigorda_photon_2021,ballantine_quantum_2021}, lattice based quantum memories~\cite{manzoni_optimization_2018} or the generation of topological phases of matter~\cite{perczel_topological_2020}.

\begin{figure}
\centering
\includegraphics[width=0.48\textwidth]{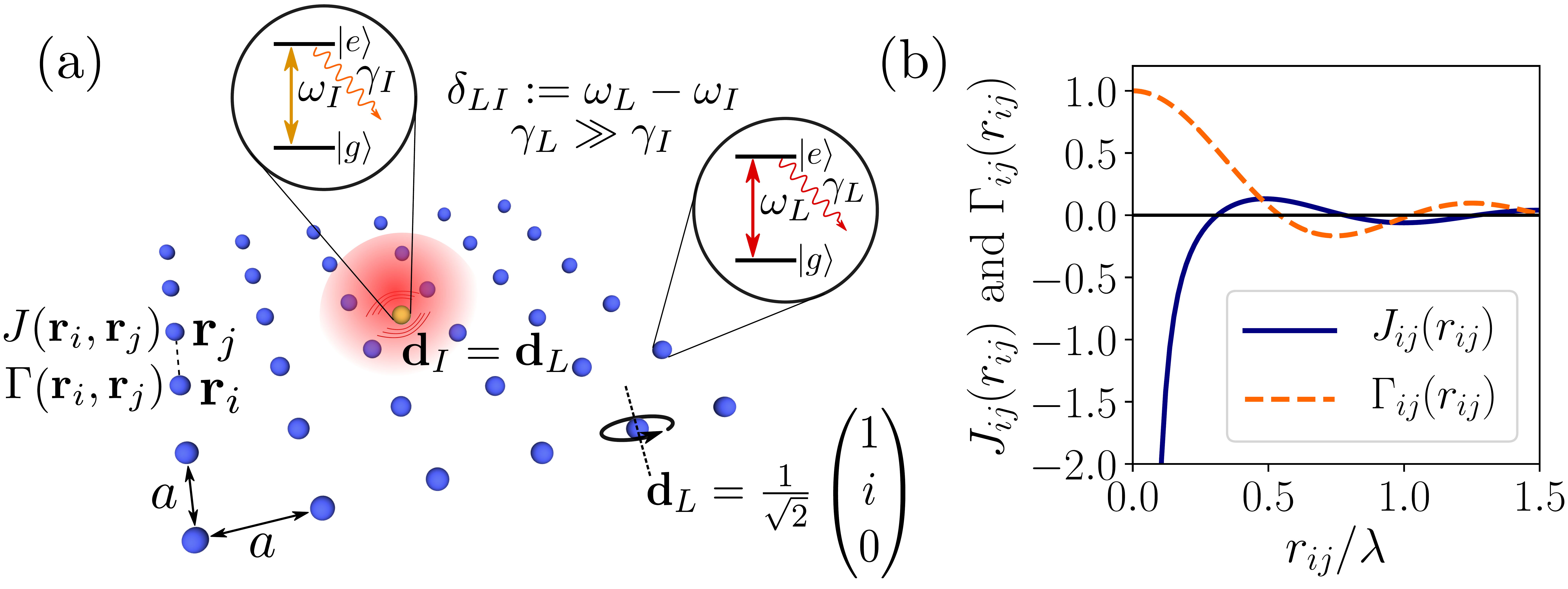}
\caption{(a) Setup -- an impurity is embedded in a periodic two-dimensional array. We analyze the effect of the lattice geometry on the effective impurity decay rate. Here, we show the interstitial case. The subsitutional case corresponds to replacing a lattice atom with an impurity. (b) Distance dependence of the coherent dipole-dipole interactions $J_{ij}(r_{ij})$ and light-induced collective dissipation $\Gamma_{ij}(r_{ij})$ as given in~\eref{eqn:couplings_g_dep}.}
\label{fig:setup}
\end{figure}
Here, we consider a setup where the collective lattice modes modify the radiative environment of an atomic impurity, see~\fref{fig:setup}(a). The impurity can either be a different atomic species compared to the lattice atoms or a different transition for the same atomic species. It was recently shown that two dimensional square arrays can act as structured Markovian baths for the impurity, effectively suppressing its decay rate by several orders of magnitude for an optimal detuning between the impurity's and the lattice atoms' transition frequency \cite{patti_controlling_2021}. The enhanced excited state lifetime also depends on the structure of the lattice. In this work, we analyze the fundamental role of the lattice geometry for the photon storage efficiency. Specifically, we focus on three key aspects: i) the performance of various lattice geometries, especially the non-centered Bravais lattices and an additional honeycomb lattice, ii) the differences between interstitial and substitutional impurity positions,~\ie between placing the impurity inside the lattice plaquette and substituting a lattice atom by an impurity, and iii) the optimal interstitial position of the impurity inside a lattice plaquette and the effect of imperfect impurity placement.

We demonstrate that the nearest neighbors of the impurity play a central role in determining the impurity's decay properties. In particular, the number of nearest neighbors and the number of distances between nearest neighbor atoms decide which geometries present a lower effective decay rate. We further find that interstitial impurity placement always results in a smaller decay rate than the substitutional case. While the minimum decay rate is found well beyond the band edge of the collective spin wave modes of the lattice for interstitial impurities, the optimal detuning for substitutional impurities is located at the band edge. This compromises the performance of substitutional impurities, as they can strongly couple to resonant lattice modes when operating at the optimal detuning. If one allows the interstitial impurity to take any position within the lattice plaquette, the optimal impurity placement corresponds to the most symmetrical points inside the plaquette, which interestingly do not correspond to the plaquette center for some of the studied geometries. Lastly, we investigate the sensitivity of decay rates to perturbations of the impurity away from its optimal position, to judge how precise experimental realizations of such atomic arrays must be to achieve suppressed decay rates. 
\section{Model}
We consider a two-dimensional lattice of quantum emitters, which interact via light-induced resonant dipole-dipole interactions. An additional impurity is placed either interstitially in the lattice plane [see~\fref{fig:setup}(a)] or at a lattice position by replacing an array atom (substitutional case). The emitters are assumed to be two-level systems with a ground state $\ket{g}$ and an excited state $\ket{e}$. The transition frequencies of the lattice atoms and the impurity are $\omega_L = 2 \pi c/ \lambda_L$ and $\omega_I = 2 \pi c/ \lambda_I$, respectively, such that the transition wavelength $\lambda_L$ is of the order of the lattice spacing. In this case, pairwise resonant dipole-dipole interactions result in collective couplings $J_{ij}(\rr_i,\rr_j)$ and collective decay rates $\Gamma_{ij}(\rr_i,\rr_j)$ for the emitters $i$ and $j$ located at positions $\rr_i$ and $\rr_j$~\cite{lehmberg_radiation_1970,lehmberg_radiation_1970_2},
\begin{subequations}
    \begin{align}
    J_{ij}(\rr_i,\rr_j) &= -\frac{3\pi \sqrt{\gamma_i \gamma_j}}{\omega} {\mathbf{d}_i^\dagger} \cdot \re [\textbf{G}(\textbf{r}_{ij}, \omega)] \cdot \mathbf{d}_j,
    \label{eqn:J} \\
    \Gamma_{ij}(\rr_i,\rr_j) &= \frac{6\pi \sqrt{\gamma_i \gamma_j}}{\omega} {\mathbf{d}_i^\dagger} \cdot \im [\textbf{G}(\textbf{r}_{ij},\omega)] \cdot \mathbf{d}_j.
    \label{eqn:Gamma} 
    \end{align}
    \label{eqn:couplings}%
\end{subequations}
Here $\gamma_{i,j}$ is the decay rate of the individual atoms $i$ and $j$, $\mathbf{d}_{i,j}$ are the respective atomic dipole moments, $\rr_{ij} = \rr_i - \rr_j$ is the vector connecting both atoms. The atoms are assumed to be point particles, which is a good approximation if the trap frequency is large enough~\cite{footnote_point_like}. We also assumed $\omega_I \approx \omega_L \equiv \omega$ in~\eref{eqn:couplings}. The couplings in~\eref{eqn:couplings} are governed by the Green's tensor for a point dipole in vacuum $\textbf{G}(\textbf{r}, \omega)$ ith components
\begin{multline} 
    G_{\alpha\beta} (\textbf{r}, \omega) = \frac{e^{i\omega r}}{4\pi r} \left[ \left( 1 + \frac{i}{\omega r} - \frac{1}{\omega^2 r^2} \right) \delta_{\alpha\beta}
    \right. \\ \left. 
    - \left( 1 + \frac{3i}{\omega r} - \frac{3}{\omega^2 r^2} \right) \frac{r_\alpha r_\beta}{r^2} \right] - \frac{\delta(\textbf{r})}{3\omega^2} \delta_{\alpha\beta},
    \label{eqn:Green}
\end{multline}
where $r=|\rr|$ denotes the distance from the dipole and $\alpha,\beta = x,y,z$. In this work, we assume both the lattice atoms and the impurity to be circularly polarized $ \mathbf{d}_L = \mathbf{d}_I = \frac{1}{\sqrt{2}}(1, i, 0)^T$. Then, the collective shifts and decay rates in~\eref{eqn:couplings} are independent of dipole orientation, and are determined solely by the distance $r_{ij}=|\rr_{ij}|$ between emitters. They can be written as 
\begin{subequations}
    \begin{align}
    J_{ij}(r_{ij}) &= -\frac{3\sqrt{\gamma_i \gamma_j}}{8 \omega r_{ij}} \left( \cos (\omega r_{ij}) + \frac{\sin (\omega r_{ij})}{\omega r_{ij}} + \frac{\cos (\omega r_{ij})}{(\omega r_{ij})^2} \right), \\
    \Gamma_{ij}(r_{ij}) &= \frac{3\sqrt{\gamma_i \gamma_j}}{4\omega r_{ij}} \left( \sin(\omega r_{ij}) - \frac{\cos(\omega r_{ij})}{\omega r_{ij}} + \frac{\sin(\omega r_{ij})}{(\omega r_{ij})^2} \right),
    \end{align}
    \label{eqn:couplings_g_dep}%
\end{subequations}
and their functional dependence is shown in~\fref{fig:setup}(b).

In the single excitation subspace the system is described by the non-Hermitian Hamiltonian $H = H_L + H_{I} + H_{LI}$. In this expression, $H_{I} = \left( \omega_I - \frac{i}{2} \gamma_I \right) s^\dag s$ is the bare Hamiltonian of the impurity, where $s=\ket{g_I}\bra{e_I}$ denotes its transition operator, $\omega_I$ its transition frequency  and $\gamma_I$ its decay rate. Note that we take $\hbar = 1$ for the remainder of this work. $H_L$ corresponds to the Hamiltonian describing the lattice atoms and $H_{LI}$ describes the interaction between the array atoms and the impurity. They are defined as
\begin{subequations}
    \begin{align}   
        H_L &= \sum_{i=1}^{N_L} \left( \omega_L - \frac{i}{2} \gamma_L \right) \sigma_i^\dag \sigma_i + \sum_{i,j \neq i}^{N_L} \left( J_{ij} - \frac{i}{2} \Gamma_{ij} \right) \sigma_i^\dag \sigma_j, \\
        H_{LI} &= \sum_{i=1}^{N_L} \left[ \left( J_{is} - \frac{i}{2} \Gamma_{is} \right) \sigma_i^\dag s + \left( J_{si} - \frac{i}{2} \Gamma_{si} \right) s^\dag \sigma_i \right],
    \end{align}
    \label{eqn:Hamiltonian}%
\end{subequations}
where $N_L$ is the number of lattice atoms and $\sigma_i=\ket{g_i}\bra{e_i}$ is the transition operator for lattice atom $i$. To simplify notation, we retain from including the argument $r_{ij}$ in the terms $J_{ij}$ and $\Gamma_{ij}$.

To quantify the photon storage efficiency of the considered setup, we calculate the effective impurity decay rate. Previous work (see Ref.~\cite{patti_controlling_2021}) presented this calculation based on the collective lattice bands in momentum space, which are found by applying Bloch's theorem for the periodic lattice. Here, we follow an alternative method to eliminate the lattice dynamics and calculate the effective impurity decay rate, derived solely in real space. In the single excitation manifold the atomic wave function can be written as $\ket{\psi(t)} = a(t)\ket{G,g_s} + \sum_{i=1}^{N_L} b_i(t)e^{i\omega_I t} \ket{e_i,g_s} + c(t) e^{i\omega_I t} \ket{G,e_s}$, where $\ket{G,g_s}$ denotes the state with all dipoles in the ground state, $\ket{e_i,g_s}$  the state where only the $i$th lattice atom is excited and $\ket{G,e_s}$ the state where only the impurity is excited. The Schrödinger equation $i \partial_t \ket{\psi(t)} = H \ket{\psi(t)}$ then results in a set of coupled equations for the amplitudes $b_i(t)$ and $c(t)$,
\begin{subequations}
\begin{align}
    \partial_t b_i(t) &= i b_i(t)\left(\delta_{LI} + \frac{i}{2} \gamma_L\right) - i \sum_{j\neq i} b_j(t)\left(J_{ij}-\frac{i}{2}\Gamma_{ij}\right)\nonumber\\
    & \quad - i c(t) \left(J_{is}-\frac{i}{2}\Gamma_{is}\right),\label{eqn:bi_eq}\\
    \partial_t c(t) &= -i \sum_{i} b_i(t)\left(J_{is}-\frac{i}{2}\Gamma_{is}\right) - \frac{\gamma_{I} }{2} c(t).
\end{align}
\end{subequations}
We introduced the detuning between the lattice and impurity atom transition frequencies as $\delta_{LI} \coloneqq \omega_I - \omega_L$. Note that we neglect any classical driving terms in the Hamiltonian in~\eref{eqn:Hamiltonian} so the derivatives of the excited state populations don't depend on the ground state population $a(t)$. Instead we will assume that the system is prepared with an excited impurity at $t=0$. The remaining set of equations can be written in Matrix form
\begin{equation} i
    \begin{pmatrix}
    \dot{b}_1(t) \\ \vdots \\ \dot{b}_{N_L}(t) \\ \dot{c}(t)
    \end{pmatrix} = \begin{pmatrix}
        ~ & ~ & ~ &   c_{1s} \\ 
        ~ & \mathbf{H}_L & ~ & \vdots \\
        ~ & ~ & ~ & c_{N_L s} & \\
        c_{s1} & \cdots & c_{s N_L} & i\gamma_I/2  
    \end{pmatrix}
    \cdot
    \begin{pmatrix}
        b_1(t) \\ \vdots \\ b_{N_L}(t) \\ c(t)
    \end{pmatrix} 
    \label{eqn:Seq_matrix}
\end{equation}
where the $N_L\times N_L$ matrix $\mathbf{H}_L$ represents the bare lattice Hamiltonian matrix containing the terms $\propto(\delta_{LI} -i\gamma_L/2)$ in the diagonal and the coupling terms $\propto \left(J_{ij} - i\Gamma_{ij}/2\right)$ in the off-diagonals [see first line in~\eref{eqn:bi_eq}]. The complex numbers $c_{i s} = c_{s i}$ represent the coupling terms between the lattice atoms and the impurity $\propto J_{is} - i\Gamma_{is}/2$.
\begin{figure*}
    \centering
    \includegraphics[width=0.95\textwidth]{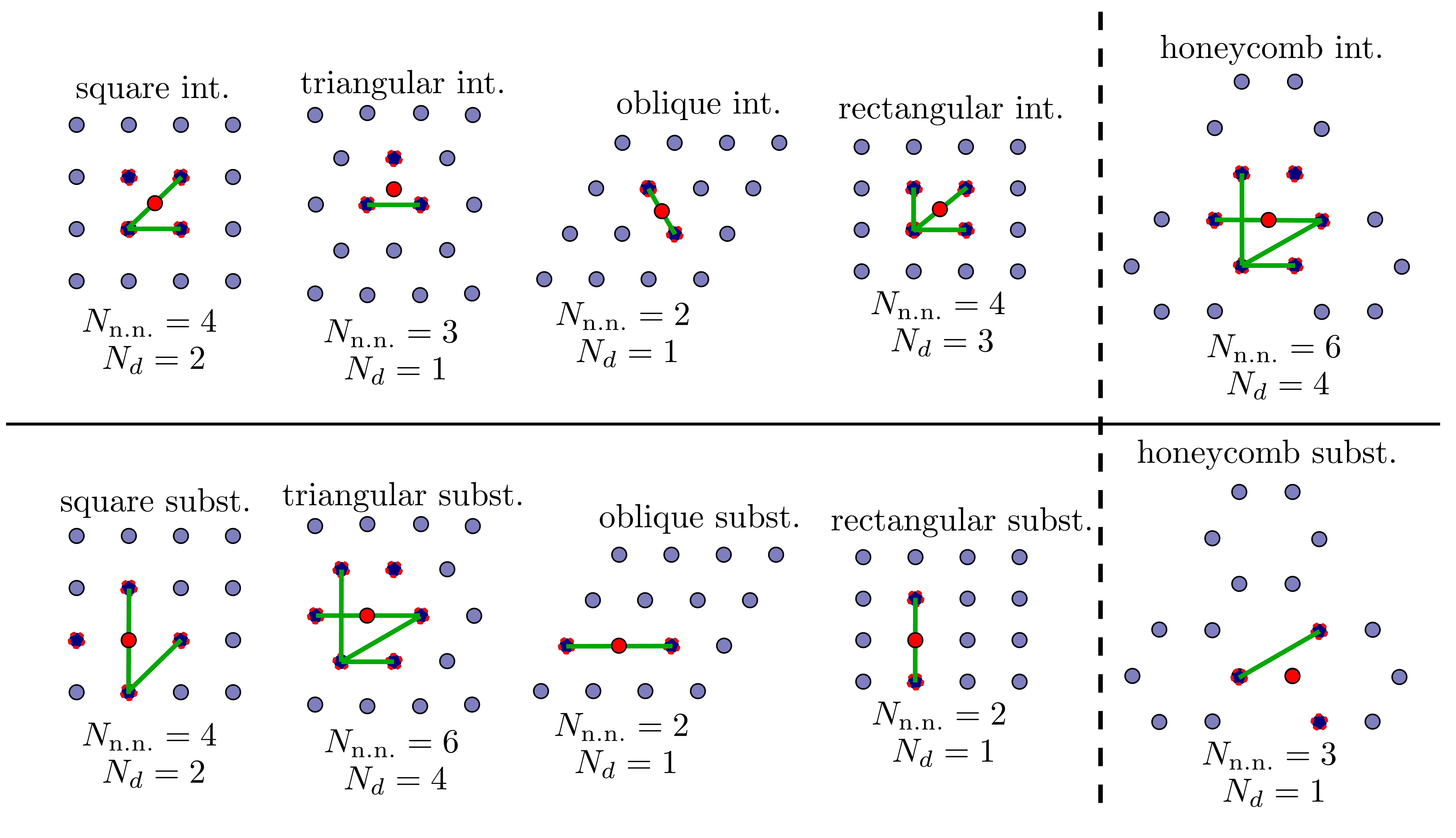}
    \caption{All lattices and cases considered in this work. The upper line corresponds to the interstitial cases whereas the second line shows the substitutional cases. The first four lattices left of the dashed line (square, triangular, oblique and rectangular) are Bravais lattices, whereas the honeycomb lattice is not a Bravais lattice. The red dot with solid black border marks the impurity and the blue dots with a dashed red boundary mark the nearest neighbors. Green lines indicate the different distances nearest neighbors can have in the particular geometries. The number $N_\mathrm{n.n.}$ specifies the number of nearest neighbors and $N_d$ the number of nearest neighbor distances for each case.}
    \label{fig:lattice_sketch}
\end{figure*}

\begin{figure}
    \centering
    \includegraphics[width=0.49\textwidth]{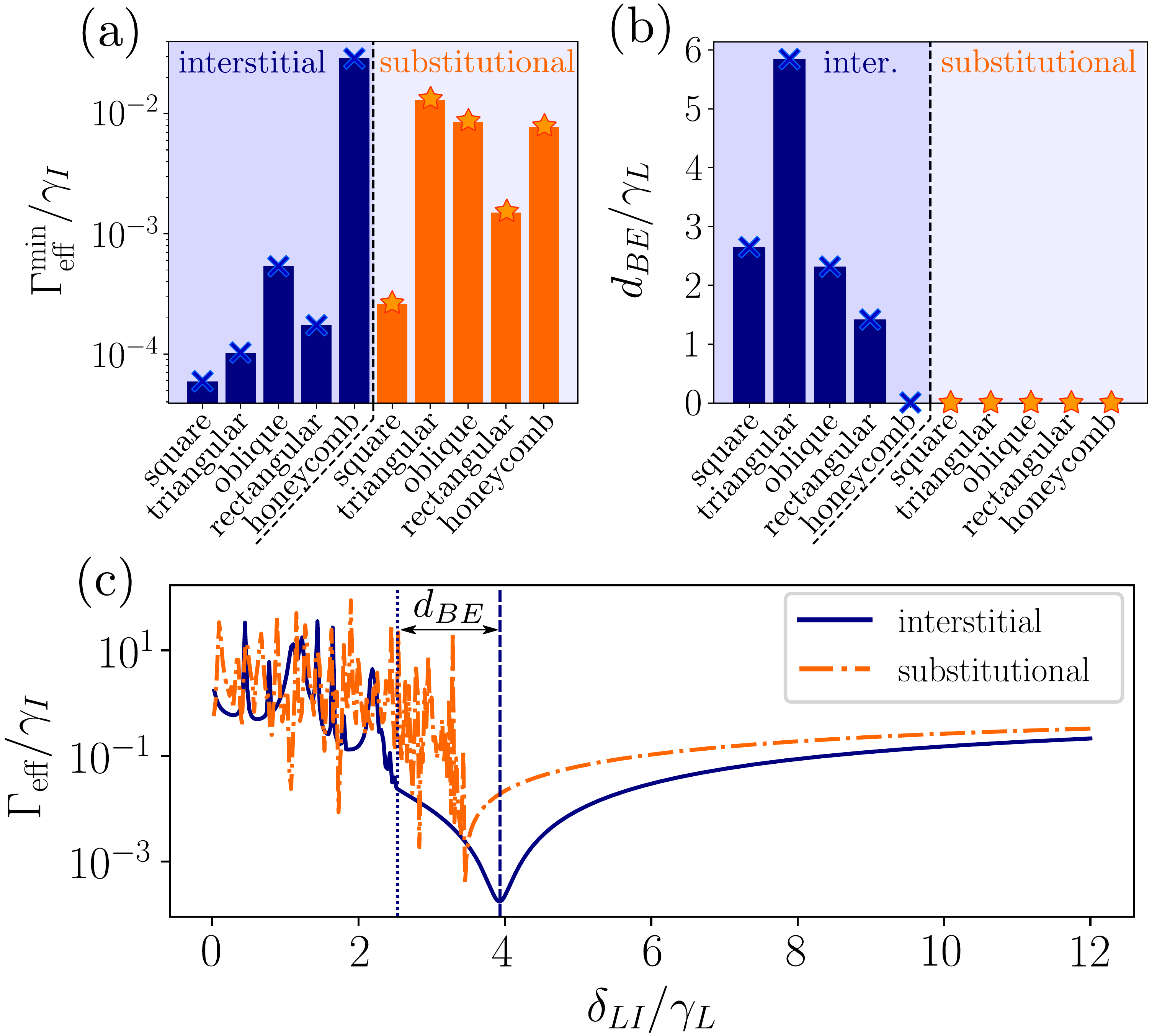}
    \caption{(a) Comparison of the minimal effective impurity decay rate $\Gamma_\mathrm{eff}^\mathrm{min}$ for the different lattice geometries considered in this work. The blue bars (with crosses indicating the maxima) correspond to interstitial impurity placement and the orange bars (with stars indicating the maxima) to the substitutional case for the corresponding lattices. (b) Band edge distance $d_{BE}$ of the lattice detuning at which the impurity decay rate is minimal. $d_{BE}$ is non-zero for all interstitial Bravais lattices and zero otherwise. (c) Exemplary curves (for the rectangular lattice case) for the effective impurity decay rate as a function of lattice-impurity detuning $\delta_{LI}$ for the interstitial (blue solid) and the substitutional (orange dash dotted) case. $d_{BE}$ marks the distance to the band edge plotted in panel (b) for all cases. It is non-zero in the interstitial case. All lattices are chosen such that the distance between the impurity and the nearest neighbor is constant when compared to the square lattice with lattice spacing $a_\mathrm{sq}=0.15\lambda$;~\cf~\eref{eqn:lattice_sp}. Other parameters: oblique lattice: $\theta = 0.3\pi$, rectangular lattice: $s = 1.5$.}
    \label{fig:g_eff_comp}
\end{figure}

If the impurity decay rate is much smaller than the lattice atoms' decay ($\gamma_I \ll \gamma_L$), the lattice acts as a Markovian bath coupled to the impurity and the lattice dynamics can be adiabatically eliminated. Defining the quantities $\mathbf{b}(t)\coloneqq (b_1(t) \hdots b_{N_L}(t))^T$ and the lattice-impurity coupling vector $\mathbf{C}_{LI}\coloneqq (c_{1s} \hdots c_{N_L s})^T$
and setting $\dot{b}_i(t) = 0$ results in the steady state for the lattice atoms,
\begin{equation}
\mathbf{b}_{ss}(t) = - \left(\mathbf{H}_L^{-1} \cdot \mathbf{C}_{LI}\right) c(t).
\end{equation}
Plugging this result back into~\eref{eqn:Seq_matrix}, we obtain the equation of motion for the impurity population $c(t)$
\begin{equation}
    \dot{c}(t) = -i \left[\frac{i}{2}\gamma_I - \mathbf{C}_{IL}^T \cdot \mathbf{H}_L^{-1}\cdot\mathbf{C}_{LI}\right] c(t),
    \label{eqn:c_dynamics}
\end{equation}
with $\mathbf{C}_{IL}^T \coloneqq (c_{s1},\hdots,c_{sN_L})$.~\eref{eqn:c_dynamics} shows that the impurity's resonance frequency and decay rate are modified by the self-energy
\begin{equation}
\Sigma_I \coloneqq -\mathbf{C}_{IL}^T \cdot \mathbf{H}_L^{-1} \cdot \mathbf{C}_{LI},
\end{equation}
which describes how the impurity is influenced by its own presence in the lattice. This allows us to define the effective decay rate for the impurity $\Gamma_\text{eff}$ as
\begin{equation}
    \Gamma_\mathrm{eff} = \gamma_I - 2~\im[\Sigma_I].
    \label{eqn:g_eff}
\end{equation}
We see that the effective impurity decay rate can be lower than the free space decay rate $\gamma_I$ if $\im[\Sigma_I]>0$. The effective decay rate $\Gamma_\mathrm{eff}$ is the central parameter for comparing different lattice geometries and impurity placements throughout this work. Note that the ultimate value of the effective decay rate non-trivially depends on the relative detuning between the lattice atoms and the impurity $\delta_{LI}$ [see also~\fref{fig:g_eff_comp}(c)].

\section{Comparison of different lattice geometries}

While a square lattice geometry is the natural choice for traditional optical lattice experiments, the recent advent of optical tweezer arrays for individual atoms~\cite{barredo_an_2016, endres_atom_2016, bernien_probing_2017, barredo_synthetic_2018, ebadi_quantum_2021}, establishes a versatile tool to realize arbitrary lattice geometries. This motivates a more general study going beyond square lattices. Current state of the art tweezer arrays don't necessarily operate in the regime where strong light-induced dipole-dipole interactions occur. However, new advances in generating optical lattices with more arbitrary geometries~\cite{struck_quantum_2011, uehlinger_artificial_2013}, and tweezer arrays of alkaline-earth atoms or lanthanides open the door to the experimental realization of this regime in the near future~\cite{olmos_long-range_2013,norcia_microscopic_2018, cooper_alkaline-earth_2018}. For example, trapping strontium atoms in a blue-detuned magic wavelength optical lattice with a lattice spacing $a=206.4\text{nm}$ and using the $^3 P_0$ $\leftrightarrow ^3 \!\!D_1$ transition at $2.6\mu\text{m}$~\cite{zhang_simulation_2021} results in $a_\mathrm{Sr}/\lambda = 0.079$. Alternatively, recent progress on cooling and trapping erbium in optical lattices would allow the generation of lattice spacings on the order of $250\text{nm}$. Using the available $1.2\mu\text{m}$ transition~\cite{ban_laser_2005} then results in $a_\mathrm{Er}/\lambda=0.2$. Hence, the lattice spacings used below are expected to be achievable in the near term. It should also be noted that the qualitative picture of the effects presented below will also be observable for larger lattice spacings as long as the condition $a<\lambda_L$ is fulfilled.

In the following, we analyze whether alternative geometries could enhance the photon storage time,~\ie diminish $\Gamma_\mathrm{eff}$ further than the square lattice geometry. To this end, we compare all four non-centered Bravais lattices,~\ie square, triangular, oblique and rectangular lattices. In addition, we also consider a honeycomb lattice to get an understanding how the photon storage efficiency behaves for non-Bravais lattices. In~\fref{fig:lattice_sketch}, we sketch all considered geometries, with interstitial cases on the top line, and substitutional cases on the bottom. The impurity atom is marked with a red dot. We also indicate the impurity's nearest neighbors (blue filled circles with red borders) and the different distances between those nearest neighbors (green lines). We analyze the difference between interstitial impurity placement (positioning the impurity in the center of the lattice plaquette), and substitutional impurity placement (replacing a lattice atom with an impurity atom). The latter is particularly relevant for optical lattice based setups, where adding an additional trap for the impurity atom at interstitial positions is challenging. The considered Bravais lattices are spanned by the following lattice vectors: i) square lattice $\mathbf{a}_1^\mathrm{sq.} = (a_\mathrm{sq.},0,0)^T$, $\mathbf{a}_2^\mathrm{sq.} = (0,a_\mathrm{sq.},0)^T$, ii) triangular lattice $\mathbf{a}_1^\mathrm{tri.} = (a_\mathrm{tri.},0,0)^T$, $\mathbf{a}_2^\mathrm{tri.} = (a_\mathrm{tri.}/2, \sqrt{3}a_\mathrm{tri.}/2,0)^T$, iii) oblique lattice $\mathbf{a}_1^\mathrm{obl.} = (a_\mathrm{obl.}, 0,0)^T$, $\mathbf{a}_2^\mathrm{obl.} = (\cot(\theta) a_\mathrm{obl.}, a_\mathrm{obl.},0)^T$ with opening angle $\theta\in(0,\pi)$ and iv) rectangular lattice  $\mathbf{a}_1^\mathrm{rec.} = (a_\mathrm{rec.} s, 0,0)^T$, $\mathbf{a}_2^\mathrm{rec.} = (0, a_\mathrm{rec}, 0)^T$ with a scaling factor $s\in(0,\infty)$. To make the geometries comparable we keep the total number of lattice atoms constant at $N_\mathrm{tot} =100$ and choose the lattice spacing such that the distance between the impurity atom and its nearest neighbors is constant.  This implies the following rescaled lattice spacings for the interstitial triangular, oblique, rectangular and honeycomb lattices if the square lattice with lattice spacing $a_\mathrm{sq.}$ is chosen as a reference:
\begin{align}
a_\mathrm{tri.} &= \frac{\sqrt{2} a_\mathrm{sq.}}{1+\tan^2(\pi/6)}, \quad 
a_\mathrm{obl.} = \frac{\sqrt{2} a_\mathrm{sq.}}{\sqrt{1-2\cot{\theta}+\cos{\theta}^2}},\nonumber\\
a_\mathrm{rec.} &= \frac{\sqrt{2}a_\mathrm{sq.}}{\sqrt{1+s^2}}, \quad \quad \, \, \, \,
a_\mathrm{honey.} = \frac{\sqrt{2}a_\mathrm{sq.}}{2}.
\label{eqn:lattice_sp}
\end{align}
For the data shown in this section, we choose a $10\times 10$ square lattice with a lattice spacing $a = 0.15\lambda$ as reference lattice. 
Note that given the fast decay with distance of the dipole-dipole coupling, nearest neighbor atoms have the greatest influence on cooperative photon storage [see~\fref{fig:setup}(b)]. However, in practice a certain lattice size is crucial because the impurity atom will be excited by a laser beam impinging onto the lattice, which as a result should be larger than the laser beam's waist. 

\fref{fig:g_eff_comp}(a) shows the minimum $\Gamma_\mathrm{eff}$ obtained at the optimal lattice-impurity detuning $\delta_{LI}$. The results for the interstitial cases are shown in blue and the subsitutional cases in orange. We find that the interstitial impurity placement always results in a smaller (and hence better) $\Gamma_\mathrm{eff}^\mathrm{min}$ compared to the substitutional case for each lattice geometry, and that the square lattice geometry always yields the smallest  $\Gamma_\mathrm{eff}^\mathrm{min}$ overall. 
When analyzing a single $\Gamma_\mathrm{eff}$ vs. $\delta_{LI}$ curve (see~\fref{fig:g_eff_comp}(c) as an example for the rectangular lattice case), another major advantage of the interstitial configuration for Bravais lattices is found.
The distance $d_{BE}$ between the detuning at which the minimal value of $\Gamma_\mathrm{eff}$ is obtained and the band edge of collective lattice modes is non-zero for interstitial impurity placement but zero for substitutional impurity placement. Note that, because we restrict our analysis to finite lattices in real-space, the notion of lattice bands is somewhat ambiguous, since we cannot define a complete momentum space basis without assuming an infinite lattice. We determine the band edge as the frequency $\delta_{LI}^{BE}$ at which strong resonances occur in the $\Gamma_\mathrm{eff}$ vs. $\delta_{LI}$ curve [see~\fref{fig:g_eff_comp}(c)]. In this regime the excited impurity resonantly couples to collective spin-waves of the lattice. Hence, the Markovian condition, which was employed to arrive at~\eref{eqn:g_eff} is no longer fulfilled,~\ie the dynamics in this regime is non-Markovian.
In~\fref{fig:g_eff_comp}(b), we plot the distance $d_{BE}$ for all considered lattice configurations. It is non-zero for interstitial impurity placement in a Bravais lattice, but zero in all other cases. In particular, it is zero for the honeycomb lattice independent of interstitial or substitutional impurity placement.
This implies that, for interstitial Bravais lattices, a well defined minimum at a finite distance from any resonant lattice mode exists [see blue curve in~\fref{fig:g_eff_comp}(c)]. Hence, the system can be easily prepared in this regime of minimal effective decay rate and it will be robust to small fluctuations in frequency. In the substitutional case, however, the optimal point -- which lies at the band edge -- is highly susceptible to tiny frequency fluctuations, which can cause resonant coupling to lattice modes and therefore diminish the photon storage properties of the impurity. Substitutional impurities should therefore be operated at a detuning slightly larger than the optimal one, which still allows to attain an enhancement of its lifetime by one or two orders of magnitude compared to free space.

\begin{figure}
    \centering
    \includegraphics[width = 0.4\textwidth]{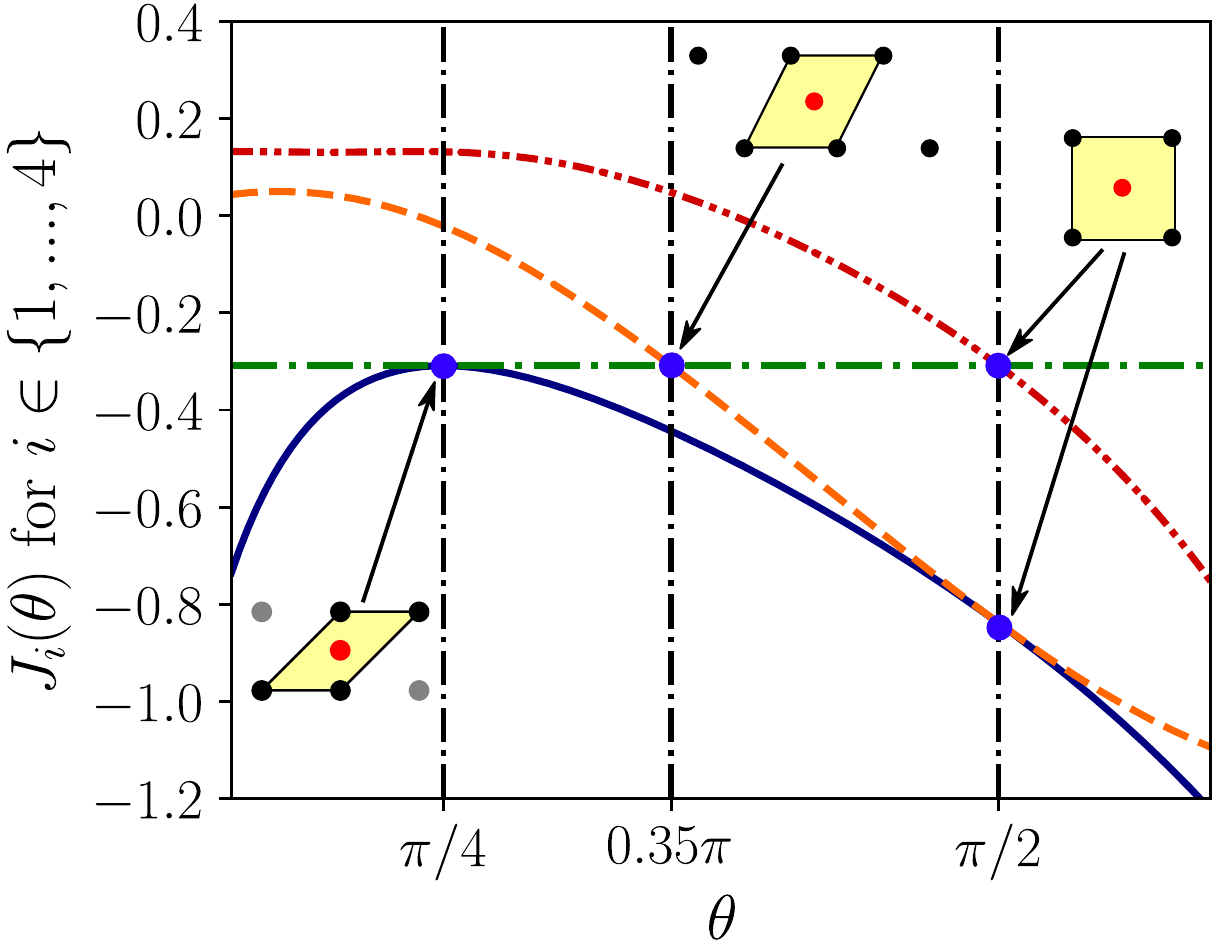}
    \caption{Functional dependence of the four coherent couplings between atoms making up a plaquette as a function of $\theta$ for an oblique lattice. The highest symmetry in couplings is obtained for $\theta = \pi/2$, which corresponds to a square lattice.
    }
    \label{fig:J_Gamma_curves}
\end{figure}

Based on the results above and by comparing the number of nearest neighbors and nearest neighbor distances shown in~\fref{fig:lattice_sketch}, we determine the following condition for optimal photon storage: A lattice geometry is optimal (smallest $\Gamma_\mathrm{eff}^\mathrm{min}$) if the impurity has the maximum number of nearest neighbors while \emph{simultaneously} the number of distances between those nearest neighbors is minimal. This condition holds for both cases interstitial and substitutional.

The condition for optimal geometries formulated above can also be understood from a slightly different angle. Ultimately, defining a lattice geometry corresponds to choosing different distances between atoms, hence, sampling a finite number of points from the dipole-dipole coupling curves shown in~\fref{fig:setup}(b). The Hamiltonian in~\eref{eqn:Hamiltonian} contains a sum over all these possible couplings among lattice atoms. Our results suggest that this sum should contain as few terms as possible to optimize the photon storage for different geometries. This symmetry in the dipole-dipole coupling terms can also be visualized when considering the transition from a square lattice to an oblique lattice. In~\fref{fig:J_Gamma_curves} we show how the coherent and dissipative dipole-dipole interactions change when transitioning from a square lattice to an oblique lattice as a function of the opening angle $\theta$ determining the oblique lattice. Under the condition that the distance between the impurity and its nearest neighbors is kept constant, the four fundamental distances between atoms defining one lattice plaquette are parametrized as a function of $\theta$ via
\begin{subequations}
\begin{align}
d_1 &= \frac{\sqrt{2}a_\mathrm{sq.}}{\sqrt{1-2\cot{\theta}+\csc^2\theta}},\\
d_2 &= \frac{\sqrt{2}a_\mathrm{sq.} \sqrt{1+\cot^2{\theta}}}{\sqrt{1-2\cot{\theta}+\csc^2{\theta}}},\\
d_3 &= \frac{\sqrt{2}a_\mathrm{sq.} \sqrt{2+\cot^2\theta-2\sqrt{1+\cot^2\theta}\cos\theta}}{\sqrt{1-2\cot{\theta}+\csc^2{\theta}}},\\
d_3 &= \frac{\sqrt{2}a_\mathrm{sq.} \sqrt{2+\cot^2\theta+2\sqrt{1+\cot^2\theta}\cos\theta}}{\sqrt{1-2\cot{\theta}+\csc^2{\theta}}}.
\end{align}
\label{eqn:sq_oblique_dist}%
\end{subequations}
One sees from~\fref{fig:J_Gamma_curves} that the most couplings for atoms in a plaquette coincide for the most symmetric case~\ie a square lattice ($\theta = \pi/2$). This implies that the square lattice is preferable compared to oblique lattices. Similar arguments can be applied to all other studied lattice geometries.

\begin{figure}
    \centering
    \includegraphics[width=\linewidth]{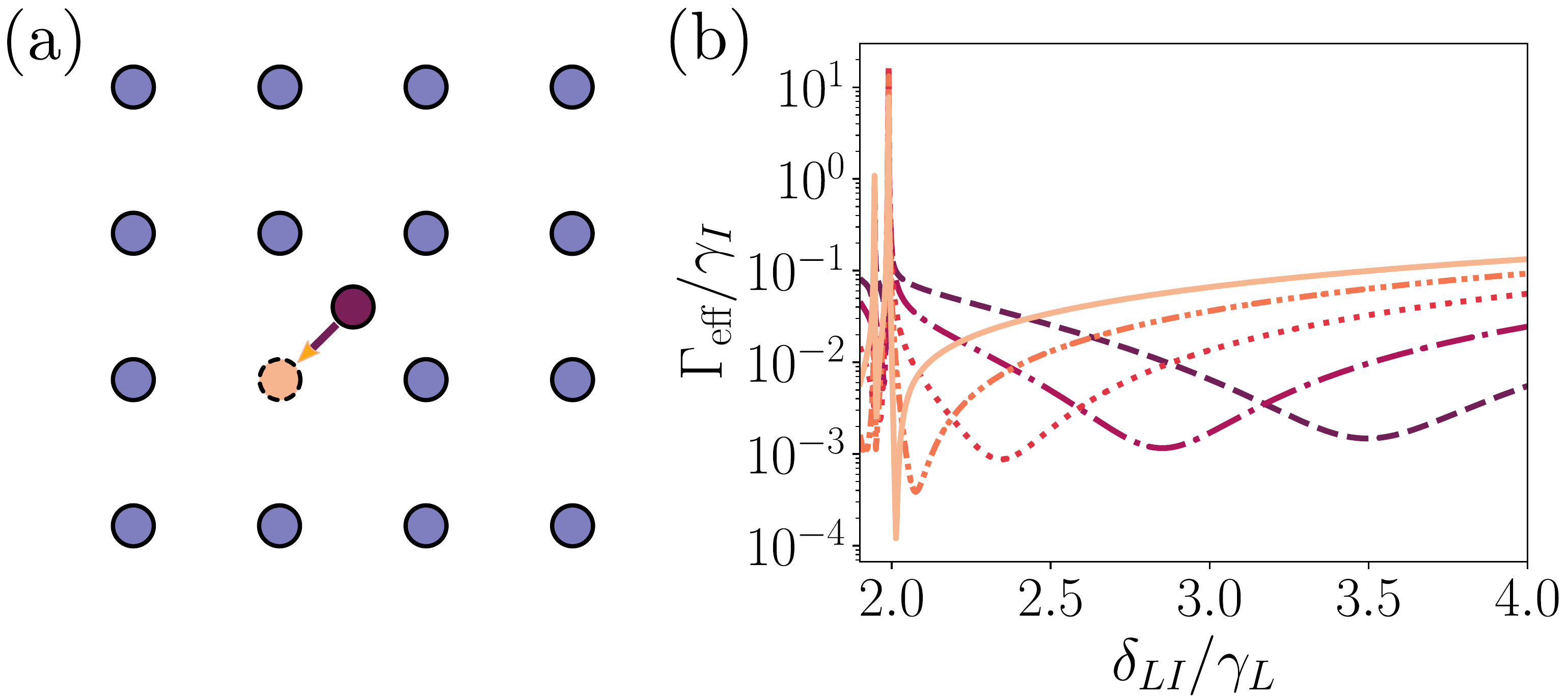}
    \caption{(a) Square lattice with a missing atom. The impurity is moved along the diagonal, from the center of the plaquette (purple with solid black border) to a lattice position (pastel orange with dashed black border), recovering then the substituational case. (b) Impurity decay rate $\Gamma_\mathrm{eff}$ as a function of lattice-impurity detuning $\delta_{LI}$ for a 10$\times$10 lattice with spacing $a=0.15\lambda$. The colorscale represents the position of the impurity. The purple dashed line corresponds to the center of the plaquette to $\mathbf{p}_1=(0,0)$, and the solid pastel orange line to the lattice position $\mathbf{p}_5=(-0.5a,-0.5a)$. The other traces correspond to $\mathbf{p}_2=(-0.1a,-0.1a)$ (dash dotted), $\mathbf{p}_3=(-0.2a,-0.2a)$ (dotted) and $\mathbf{p}_4=(-0.3a,-0.3a)$ (dash-dot-dotted).}
    \label{fig:moving_impurity}
\end{figure}

While nearest neighbors determine to a large extent the suppression of the decay rate of the impurity, quantities like the optimal detuning from the band edge depend on the symmetry of the whole lattice and the exact position of the impurity in the array. For example, the square interstitial and square substitutional lattices have the same number of nearest neighbors and nearest neighbor distances, but result in $d_{BE} \neq 0$ in the first case and $d_{BE}=0$ in the second. \fref{fig:moving_impurity} shows how this transition occurs. We first consider an interstitial square lattice with a missing nearest neighbor, which presents a minimum decay rate away from the band edge, as shown by the purple trace in~\fref{fig:moving_impurity}(b). Note that the achieved $\Gamma_\mathrm{eff}$ is larger than the one reported in~\fref{fig:g_eff_comp}(a) due to a missing nearest neighbor. If the impurity is moved along the diagonal towards the position of the missing lattice atom the optimal detuning is continuously shifted towards the band edge. When the impurity reaches the position of the missing lattice atom, we recover the substitutional case and the minimum decay rate occurs exactly at the band edge (see pastel orange trace).
This phenomenon arises from two different processes. First, the position of the band edge solely depends on the geometry of the array -- a square lattice with spacing $a$ for the case shown here -- and therefore remains constant at $\delta_{LI}=2 \gamma_L$ for a lattice with $a=0.15\lambda$. When the impurity is placed at the position of the missing atom, it lies again at the center of a square plaquette whose axes are rotated by $45^\circ$ and whose lattice constant has increased to $a'=\sqrt{2}a$. This suggests that the optimum decay rate of the substitutional lattice should be similar to that of a square interstitial lattice with spacing $a'=\sqrt{2}\times0.15$, that is $\delta_{LI} \approx 1.95 \delta_L$, which approximately lies at the band edge of the full lattice. Note that the rotated square lattice has two atoms per unit cell. It was numerically confirmed that removing the excess atom shifts the band edge to a lower detuning, but leaves the minimum decay rate at around $\delta_{LI}=2 \gamma_L$, thus recovering the interstitial case for a larger lattice spacing.

\section{Interstital case - optimal impurity position}\label{sec:opt_pos}
\begin{figure*}
    \centering
    \includegraphics[width = 0.99\textwidth]{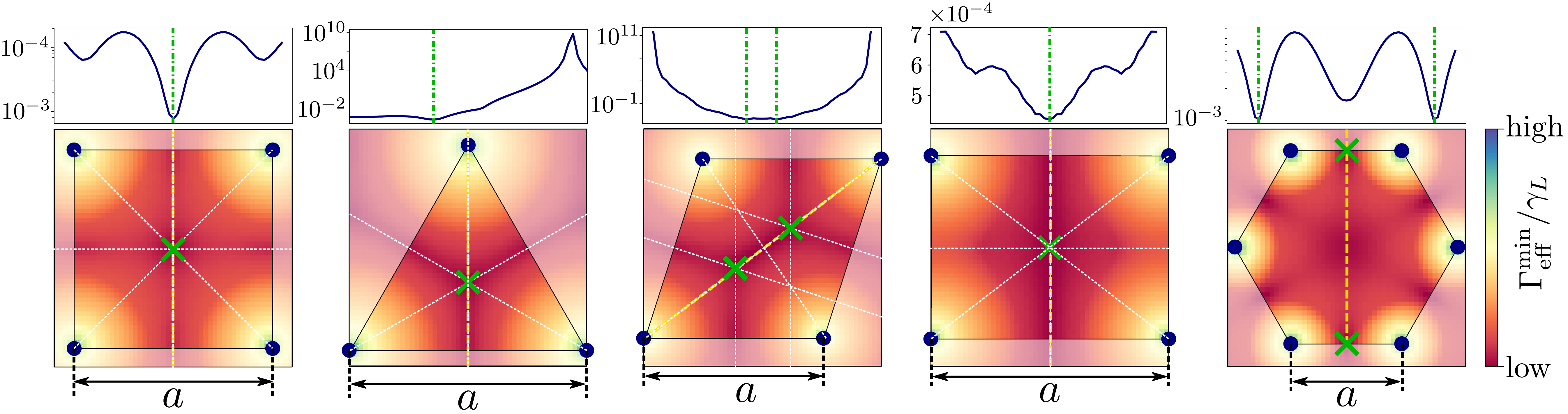}
    \caption{Optimal impurity positioning for the interstitial case for all lattices considered in this work. The color coding shows the minimal effective decay rate $\Gamma_\mathrm{eff}/\gamma_L$, which can be obtained by placing the impurity at the respective position in the plaquette. The optimal impurity position(s),~\ie the positions at which $\Gamma_\mathrm{eff}^\mathrm{min}/\gamma_L$ exhibits a global minimum, is (are) indicated by green cross(es). For the oblique and the honeycomb lattice multiple global minima away from the palquette center are found. The top panels show cuts along the orange dashed lines and the green dash-dotted lines indicate the respective global minima positions of $\Gamma_\mathrm{eff}^\mathrm{min}/\gamma_L$. The white dashed lines indicate the lines along which the distance between two plaquette atoms is always equal. These lines make up the Voronoi partition of the lattice.
    }
    \label{fig:opt_imp_pos}
\end{figure*}
In the previous section we placed the impurity in the center of the lattice plaquette for the interstitial case. Here, we analyze if this is the optimal placement of the impurity for all considered geometries. For a given impurity position, $\delta_{LI}$ can be chosen to give optimal $\Gamma_\text{eff}$ by minimizing along a similar curve as the one shown in~\fref{fig:g_eff_comp}(c). By conducting this optimization for all impurity positions within a plaquette, a map of the optimal impurity placement can be constructed. The results of this procedure are depicted in~\fref{fig:opt_imp_pos}. We observe, that the center of the plaquette is the optimal impurity position (indicated by green crosses in~\fref{fig:opt_imp_pos}) for the square, triangular and rectangular lattices, but \emph{not} for the oblique and honeycomb lattice.

In all cases, geometric symmetries determine where the points of minimal $\Gamma_\text{eff}$ lie. For all Bravais lattices, the paths of minimal $\Gamma_\text{eff}$ follow the lines along which the distances to two nearest lattice points are equal. Such partitions of a lattice are commonly referred to as Voronoi partition~\cite{aurenhammer_voronoi_1991} (see white dashed lines in~\fref{fig:opt_imp_pos}). The vertices where edges of this Voronoi partition meet are the points of minimal $\Gamma_\text{eff}$ and correspond to the global minima. The number of edges that coincide at any one point roughly correlates to how low the effective decay rate will be compared to high symmetry points of other lattices. For instance, the center of the square plaquette, where four edges coincide, has an optimal $\Gamma_\text{eff}^\mathrm{min} = 5.94\times 10^{-5}\gamma_L$ that is approximately an order of magnitude less than the optimal $\Gamma_\text{eff}$ at the center of the triangular plaquette ($1.03\times 10^{-4}\gamma_L$), where only three edges coincide.

The Voronoi partitions of the lattice plaquettes also capture the fact that the oblique lattice has two optimal positions other than the center of the plaquette. In this case the optimal positions are on the line of symmetry along the long diagonal of the plaquette, at a pair of points where three Voronoi edges meet. As we approach the square lattice at $\theta = \frac{\pi}{2}$, these two optima merge into a single one located at the center of the plaquette.

The role that symmetry plays in creating the Voronoi geometries can again be understood based on the results shown in~\fref{fig:J_Gamma_curves}. In general, the larger the number of couplings, the lower the $\Gamma_\mathrm{eff}$ for an impurity placed at that optimal point. In this way, lattices with higher degrees of symmetry (i.e. the square and triangular lattices) and with higher numbers of atoms in a single plaquette (i.e. any Bravais lattice other than the triangular lattice) possess impurity positions with the smallest $\Gamma_\text{eff}$. The fact that the square lattice possesses both of these properties helps to explain why it stands out amongst the various Bravais lattices as the optimal choice.

Note that the results presented in~\fref{fig:opt_imp_pos} also imply that even if an optimal impurity placement is not possible due to experimental constraints or imperfections, placing the impurity in the vicinity of this optimal point still allows for a significantly enhanced photon storage time compared to the free space case. This also applies to slight disorder in the positioning of the lattice atoms, as long as the disorder is much smaller than the lattice period~\cite{shahmoon_cooperative_2017}.

\section{Conclusions and Outlook}\label{sec:conclusion}
By performing a detailed analysis of different geometries we determined that the number of nearest neighbors and in particular the number of distances between nearest neighbors plays a crucial role in determining the optimal geometry for enhanced photon storage in an impurity interacting with an atomic array. While the substitutional case,~\ie substituting a lattice atom with an impurity is found to be always worse (\ie results in smaller photon storage times) compared to interstitial impurity placement, it still allows an enhancement compared to the free space case of several orders of magnitude. The dominant role of the impurity's nearest neighbors is also beneficial for potential experimental implementations because lattice vacancies will not have a large impact as long as they do not involve nearest neighbors. Note that the results concerning the fundamental role of the lattice symmetry presented in this work also hold for efficient coherent coupling of multiple impurities via collective lattice modes~\cite{masson_atomic-waveguide_2020, patti_controlling_2021}. 
While this work focuses on atom arrays, our model applies to a wide array of quantum emitters in the solid state, such as excitons in transition metal dichalcogenides (TMDs)~\cite{Palacios-Berraquero2018,scuri_large_2018,back_realization_2018} and arrays of nitrogen and silicon vacancy centers in diamond~\cite{tamura_array_2014,sipahigil_an_2016}. That being said, a complete modelling of these systems would require the inclusion of further effects, such as dephasing, emitter delocalization, and a more thorough analysis of lattice vacancies. For TMDs, the emitters can in general no longer be described as point particles and a modified model needs to be developed. Such avenues constitute but a few of the numerous potential extensions of this work.

The results on the optimal impurity positioning presented in section~\ref{sec:opt_pos} also suggest a promising future research direction. The values of $\Gamma_\mathrm{eff}^\mathrm{min}/\gamma_L$ for each impurity position render an effective potential generated by the lattice for the impurity atom. Hence, including motional degrees of freedom for the impurity~\cite{shahmoon_quantum_2020} could result in non-trivial dynamic phenomena~\cite{chang_self-organization_2013}. Note that in this work we solely focused on circular polarizations for both the lattice atoms and the impurity. In general, the atomic polarizations are another parameter to optimize for enhanced photon storage. While a detailed investigation goes beyond the scope of the present manuscript, some insights can be obtained via related works optimizing geometries for minimizing the collective light shifts for optical lattice clocks~\cite{kramer_optimized_2016}.
The lattice geometry is also expected to play a role beyond the single-excitation manifold where non-linear quantum effects such as photon blockade and entanglement can occur as was pointed out in recent works~\cite{cidrim_photon_2020, williamson_superatom_2020, bettles_quantum_2020}.

In this work we focused on periodic lattice geometries, which in general allow the extension to infinite size by proper definition of unit vectors. In general, alternative geometries such as bio-inspired coupled nano-rings~\cite{moreno-cardoner_subradiance-enhanced_2019, cremer_polarization_2020, moreno-cardoner_efficient_2022} can be studied in a similar fashion. Besides, finding more general geometries in two- or three dimensions that might enhance the photon storage times even further via tailored machine learning algorithms~\cite{jiang_free-form_2019,kudyshev_machine-learning-assisted_2020} is an exciting research avenue for the future.

\emph{Acknowledgments.} S.~O. is supported by a postdoctoral fellowship of the Max Planck Harvard Research Center for Quantum Optics. O.~R.~B. acknowledges support from Fundació Bancaria “la Caixa” (LCF/BQ/AA18/11680093). S.~F.~Y. would like to acknowledge funding from NSF (formalism), AFOSR (calculations and numerics), and DOE (potential applications to solid state surfaces).

The numerical simulations were performed with the open-source framework \texttt{QuantumOptics.jl}~\cite{kramer_quantumopticsjl_2018}.

\appendix
\section{Data for case comparison}
Here we provide the data, which were obtained by minimizing $\Gamma_\mathrm{eff}$ along $\delta_{LI}$ to generate~\fref{fig:g_eff_comp}.
\begin{table}[h]
  \centering
  \begin{tabular}{ p{2.6cm}||P{0.8cm}|P{0.8cm}|P{1.8cm}|P{1.3cm}}
  \hline
 case   & $N_{n.n.}$ & $N_d$ &  $\Gamma_\mathrm{eff}^\mathrm{min}/\gamma_L$ &$d_{BE}/\gamma_L$\\
 \hline
 square int.        & 4 & 2 & $5.94\times10^{-5}$ & 2.65\\
 triangular int.    & 3 & 1 & $1.03\times10^{-4}$ & 5.845\\
 oblique int.    & 2 & 1 & $5.38\times10^{-4}$ & 2.32\\
 rectangular int.   & 4 & 3 & $1.74\times10^{-4}$ & 1.42\\
 honeycomb int.     & 6 & 4 & $2.9\times10^{-2}$  & 0.0\\
 \hline
 \hline
 square subst.      & 4 & 2 & $2.63\times10^{-4}$ & 0.0\\
 triangular subst.  & 6 & 4 & $1.3\times10^{-2}$ & 0.0\\
 oblique subst.  & 2 & 1 & $8.54\times10^{-3}$ & 0.0\\
 rectangular subst. & 2 & 1 & $1.51\times10^{-3}$ & 0.0\\
 honeycomb subst.   & 3 & 1 & $7.78\times10^{-3}$ & 0.0\\
 \hline
\end{tabular}
  \caption{Data to generate the bar plots shown in~\fref{fig:g_eff_comp}(a) and (b).}
  \label{tab:1}
\end{table}


\begin{thebibliography}{59}
\bibitem [{foo()}]{footnote_point_like}%
  \BibitemOpen
  \href@noop {} {\bibinfo  {journal} {This is a well justified approximation to
  determine the fundamental influence of the lattice geometry. For typical
  ultracold atom experiments trap frequencies up to 100 kHz are realistic. If
  the atoms are cooled to the motional ground state of the corresponding
  harmonic trapping potential, the ground state wave function has a width
  $\sigma = \sqrt{\hbar/(m \omega)}$ of tens of nano meters. The finite width
  of the atomic wavefunction will only result in quantitative changes of
  $J_{ij}$ and $\Gamma_{ij}$}\ }\BibitemShut {NoStop}%
\end{thebibliography}

%

\end{document}